\documentclass[12pt]{article}
\usepackage{amsmath} 
\input{epsf} 
\def\naive{na\"{\i}ve} 
\newcommand\as{\alpha_{\mathrm{S}}} 
\newcommand\f[2]{\frac{#1}{#2}} 
\def\beq{\begin{equation}} 
\def\eeq{\end{equation}} 
\def\beeq{\begin{eqnarray}} 
\def\eeeq{\end{eqnarray}} 
 
\def\to{\rightarrow}
\def\ito{\leftarrow} 
\def\nn{\nonumber}

\def\ms{${\overline {\rm MS}}$}

\def\pitcut{\pi_T}

\begin{document} 
\begin{titlepage}
\begin{flushright}
ZU-TH 12/11
\end{flushright}
\renewcommand{\thefootnote}{\fnsymbol{footnote}}
\vspace*{2cm}

%\begin{center}
%{\Large \bf Transverse-momentum resummation\\[0.1cm] 
%and Higgs boson 
%production:\\[0.3cm] 
%hard-collinear coefficients at the NNLO}
%\end{center}
\begin{center}
{\Large \bf Higgs boson 
production at hadron colliders:\\[0.3cm]
%and transverse-momentum resummation:\\[0.4cm] 
hard-collinear coefficients at the NNLO}
\end{center}

\par \vspace{2mm}
\begin{center}
{\bf Stefano Catani}$^{(a)}$~~and~~{\bf Massimiliano Grazzini}$^{(b)}$\footnote{On leave of absence from INFN, Sezione di Firenze, Sesto Fiorentino, Florence, Italy.}\\

\vspace{5mm}

$^{(a)}$ INFN, Sezione di Firenze and Dipartimento di Fisica e Astronomia,\\ 
Universit\`a di Firenze,
I-50019 Sesto Fiorentino, Florence, Italy\\

$^{(b)}$ Institut f\"ur Theoretische Physik, Universit\"at Z\"urich, CH-8057 Z\"urich, Switzerland

\vspace{5mm}

\end{center}

\par \vspace{2mm}
\begin{center} {\large \bf Abstract} \end{center}
\begin{quote}
\pretolerance 10000

We consider the production of the Standard Model Higgs boson
through the gluon fusion mechanism in hadron collisions.
We present the next-to-next-to-leading order (NNLO) QCD result
of the hard-collinear coefficient function 
for the all-order resummation of logarithmically-enhanced
contributions at small transverse momentum.
The coefficient function controls NNLO contributions 
in resummed calculations at full next-to-next-to-leading logarithmic accuracy.
The same coefficient function is 
used
in applications of the subtraction method to perform 
fully-exclusive perturbative calculations up to NNLO.

\end{quote}

\vspace*{\fill}
\begin{flushleft}
June 2011
\end{flushleft}
%
% Revised Version submitted to Journal 
\end{titlepage}

\setcounter{footnote}{1}
\renewcommand{\thefootnote}{\fnsymbol{footnote}}

The transverse-momentum $(q_T)$
distribution of systems with high invariant mass $M$
(such as Drell--Yan lepton pairs, photon pairs, 
%vector boson(s), Higgs boson(s), 
vector bosons, Higgs bosons,
and so forth)
produced in hadron collisions 
is computable by using perturbative QCD.
However, in the small-$q_T$ region (roughly, in the region where
$q_T \ll M$) the convergence of the fixed-order perturbative expansion
in powers of the QCD coupling $\as$ is spoiled by 
the presence of large logarithmic 
%coefficients 
contributions
of the type $\ln^n(M^2/q_T^2)$.
The predictivity of perturbative QCD can be recovered through the summation
of these logarithmically-enhanced contributions to all order in $\as$
\cite{Dokshitzer:hw}.

The structure of the resummed calculation can be organized in a
process-independent form 
\cite{Collins:1981uk, Collins:1984kg, Catani:2000vq, Catani:2010pd},
%\cite{Collins:1981uk}--\cite{Catani:2010pd}, 
%where
in which
the logarithmic contributions
are controlled by a set of perturbative functions,
%which are
usually denoted as
$A(\as), B(\as), C(\as)$ and ${\cal H}(\as)$ 
(see, e.g., Eqs.~(\ref{whath}) and (\ref{HCCGG})
 and related comments).
These functions
%(to a large extent, they are process independent)
and, hence, their perturbative coefficients (e.g. the coefficient 
$A^{(n)}$ of the $n$-th order contribution $A^{(n)} \as^n$ to $A(\as)$),
have no explicit dependence on the ratio $q_T/M$.
%These 
The perturbative coefficients, once they are known, can be inserted in
process-independent resummation formulae that systematically resum, in explicit
form, the classes of leading, next-to-leading, next-to-next-to-leading
(and so forth) logarithmic contributions to the transverse-momentum 
distribution.
In this respect, the transverse-momentum resummation program has formal
analogies\footnote{These analogies may hide important physical, conceptual
and technical differences, which are discussed in the literature on
transverse-momentum resummation.}
with the study of logarithmic scaling violations (of ultraviolet or collinear
origin), where the resummation of logarithmic terms is traded for the
calculation
of perturbative functions, such as short-distance coefficient functions and
anomalous dimensions.

Most of the $q_T$ resummation coefficients are known, since some time
\cite{Kodaira:1981nh, Davies:1984hs, Catani:vd, 
%deFlorian:2000pr, 
deFlorian:2001zd}, up to the second order in $\as$.
The third-order coefficient $A^{(3)}$ has been obtained in 
Ref.~\cite{Becher:2010tm}.
%recently. 
In recent years, we have been working on a research project aimed at the
completion of the $q_T$ resummation program at the second perturbative order.
This requires the calculation of the second-order coefficient 
function\footnote{In this introductory part we are using a shorthand 
%language,
notation,
since the symbol ${\cal H}^{(2)}$ actually refers to a set of several
coefficient functions.}
${\cal H}^{(2)}(z)$ (see Eq.~(\ref{H2})), which includes a process-dependent
part. 
%The ${\cal H}^{(2)}$ coefficients have been explicitly computed for two
%benchmark processes, namely Higgs boson production and the Drell--Yan process,
%and the corresponding results have been announced
%and used in Refs.~\cite{Catani:2007vq} and \cite{Catani:2009sm}, respectively.
The computation of the ${\cal H}^{(2)}$ coefficients has been explicitly
carried out for two 
benchmark processes, namely, Higgs boson production and the Drell--Yan process,
and the corresponding results have been obtained
and used in Refs.~\cite{Catani:2007vq} and \cite{Catani:2009sm}, respectively.
In the case of the Drell--Yan process, the result has also been applied 
\cite{Bozzi:2010xn} to the $q_T$ spectrum of the $Z$ boson, by explicitly
performing transverse-momentum resummation at the next-to-next-to-leading
logarithmic (NNLL) accuracy.
In this 
%letter, 
paper,
we consider Higgs boson production and we document the result
for ${\cal H}^{(2)}$ \cite{Catani:2007vq} in explicit analytic form.
We also illustrate the method that we have used to perform the calculation.
%of ${\cal H}^{(2)}$.

Considering the perturbative contributions that are logarithmically-enhanced and,
thus, singular in the limit $q_T \to 0$, the calculation of ${\cal H}^{(2)}$
completes the knowledge of the $q_T$ distribution at full 
next-to-next-to-leading order (NNLO) accuracy. This fact has implications in the
context of both {\em resummed} and {\em fixed-order}  calculations,
as we briefly discuss below.
%and we
%briefly comment on that. The comments are quite general, in view of the
%process-independent features of ${\cal H}^{(2)}$.

Some recent resummed calculations of the $q_T$ spectrum of the Standard Model 
(SM) Higgs  boson at Tevatron and LHC energies are presented in 
Refs.~\cite{Balazs:2000wv}--\cite{Mantry:2010mk}.
%Refs.~\cite{Balazs:2000wv, Berger:2002ut, Bozzi:2003jy, 
%Kulesza:2003wi, Kulesza:2003wn, Bozzi:2005wk, Bozzi:2007pn, Mantry:2010mk}.
The inclusion of ${\cal H}^{(2)}$ in calculations that use the other 
$q_T$ resummation coefficients up to NNLL order gives theoretical predictions
that embody the exact NNLO calculation in the small-$q_T$ region. The 
NNLL resummed calculations can then be properly matched 
(by using, for instance, $q_T$ resummation as in the formalism of 
Ref.~\cite{Bozzi:2005wk}) with the customary fixed-order calculation at large 
$q_T$, in such a way that the integration over $q_T$ of the 
$q_T$ distribution exactly returns the NNLO value of the {\em total}
cross section. Indeed, a rough approximation of ${\cal H}^{(2)}$,
such as to reproduce the NNLO value of the {\em total} cross section
with good numerical accuracy, was constructed and used in 
Ref.~\cite{Bozzi:2005wk}. The approximation of Ref.~\cite{Bozzi:2005wk}
represents a very crude estimate of the function ${\cal H}^{(2)}(z)$;
nonetheless, that approximation quantitatively works very well 
(especially at LHC energies)
\cite{Bozzi:2005wk, Bozzi:2007pn} over a wide range of Higgs boson masses.
An updated version of the code
${\tt HqT}$ \cite{Bozzi:2005wk}, which also implements the exact coefficient 
${\cal H}^{(2)}$, 
%will be available shortly.
is now available \cite{deFlorian:2011xf}.

In Ref.~\cite{Catani:2007vq}, we presented a practical formalism to perform
NNLO calculations at the fully-exclusive level for a specific class of
processes, namely, the production of colourless
high-mass systems in hadron collisions. The formalism exploits the subtraction
method to cancel the unphysical infrared divergences that separately occur in
the real and virtual radiative corrections. The explicit construction of the
subtraction counterterms \cite{Bozzi:2005wk, Bozzi:2007pn} is based on the
process-independent structure of transverse-momentum resummation formulae and
on their expansion up to NNLO in QCD perturbation theory. The formalism thus
requires the complete knowledge of the $q_T$ resummation coefficients up to 
${\cal O}(\as^2)$. Although the results of the present paper were not
explicitly illustrated in  Ref.~\cite{Catani:2007vq}, they were taken into
account in the NNLO computations presented therein. In particular, the explicit
application to Higgs boson production (which was implemented in the 
Monte Carlo code ${\tt HNNLO}$) considered in 
Refs.~\cite{Catani:2007vq,Grazzini:2008tf} is based on and implements 
the analytic results for the coefficient function ${\cal H}^{(2)}$ that are
documented in the present paper.

The paper is organized as follows.
We first introduce our notation and describe the small-$q_T$ behaviour of the
Higgs boson cross section up to NNLO. Then we briefly review
transverse-momentum resummation for Higgs boson production  and the
corresponding all-order resummation formula recently derived in 
Ref.~\cite{Catani:2010pd}. The new resummation formula differs from its 
\naive\
version 
that is used in the literature: the differences start at ${\cal O}(\as^2)$,
which is relevant for the purposes of the present paper. Finally, we present
the analytic results of our NNLO calculation of the $q_T$ distribution.
The results are expressed directly in terms of the coefficient function
${\cal H}^{(2)}(z)$ and related resummation coefficients. We conclude the paper
by describing the method that we have used to perform the NNLO calculation.

We briefly introduce the theoretical framework and our notation.
We consider the production of the SM Higgs boson $H$,
through the gluon fusion mechanism $gg \to H$, in hadron--hadron collisions.
The effective coupling $ggH$ is produced by heavy-quark loops, and the 
top quark gives the dominant contribution. We treat the coupling
$ggH$ in the framework of the 
large-$m_{top}$ approximation 
\cite{Ellis:1975ap, Kramer:1996iq, Chetyrkin:1997iv}, 
and we consider a single heavy quark, 
the top quark with mass $m_{top}$, and $n_F$ $(n_F=5)$ 
%light-quark 
massless-quark flavours.
We use the narrow width approximation and we treat the Higgs boson as an
on-shell particle with mass $M$.
The QCD expression of the Higgs boson transverse-momentum
cross section is
\begin{equation}
\label{dcross}
\f{d\sigma}{d q_T^2}(q_T,M,s)=\sum_{a,b}
\int_0^1 dz_1 \,\int_0^1 dz_2 \,f_{a/h_1}(z_1,M^2)
\,f_{b/h_2}(z_2,M^2) \;
\f{d{\hat \sigma}_{ab}}{d q_T^2}(q_T, M,{\hat s}=z_1z_2s; \as(M^2)) 
\;,
\end{equation}
where $f_{a/h_i}(x,\mu_F^2)$ ($a=q_f,{\bar q_f},g$) are the parton densities of 
the colliding hadrons ($h_1$ and $h_2$) at the factorization scale $\mu_F$,
and $d{\hat \sigma}_{ab}/d q_T^2$ are the
partonic cross sections. The centre--of--mass energy of the two colliding
hadrons is denoted by $s$, and ${\hat s}$ is the partonic centre--of--mass
energy. We use parton
densities as defined in the \ms\
factorization scheme, and $\as(\mu_R^2)$ is the QCD running coupling 
at the renormalization scale $\mu_R$ in the \ms\
renormalization scheme. In Eq.~(\ref{dcross}) and throughout the paper,
the arbitrary factorization and renormalization scales, $\mu_F$ and $\mu_R$,
are set to be equal to the Higgs boson mass $M$.

The partonic cross sections $d{\hat \sigma}_{ab}/d q_T^2$ are computable in
QCD perturbation theory as power series expansions in $\as(M^2)$.
We are interested in the perturbative contributions that are large in the 
small-$q_T$ region $(q_T \ll M)$ and, eventually, singular in the limit 
$q_T \to 0$.
To explicitly 
%illustrate 
present
the perturbative structure of these enhanced terms at
small $q_T$, we integrate the $q_T$ distribution over the 
region $0 \leq q_T \leq Q_0$, and we introduce the cumulative partonic cross
section
\begin{equation}
\label{inte}
\int_0^{Q_0^2}dq_T^2 
\;\f{d{\hat \sigma}_{ab}}{dq_T^2}(q_T,M,{\hat s}=M^2/z;\as(M^2))
\equiv z \,\sigma_H^{(0)}(\as(M^2)) 
\;{\hat R}_{ab}(z,M/Q_0;\as(M^2)) \;\;,
\end{equation}
where the overall normalization of the function ${\hat R}_{ab}$ is defined with
respect to $\sigma_H^{(0)}$,
which is the Born level cross section for the partonic subprocess $gg\to H$.
Using the large-$m_{top}$ approximation, the explicit expression of 
$\sigma_H^{(0)}$ is \cite{Ellis:1975ap}
\begin{equation}
\label{sig0}
\sigma_H^{(0)}(\as)=\f{G_F \;\as^2}{288 \,\pi \,\sqrt{2}} \;\;,
\end{equation}
where $G_F$ is the Fermi constant.
The partonic function ${\hat R}$ has the following perturbative expansion
\begin{equation}
\label{eqnR}
{\hat R}_{ab}(z,M/Q_0;\as)=\delta_{ga} \,\delta_{gb} \,\delta(1-z)+
\sum_{n=1}^\infty
\left(\f{\as}{\pi}\right)^n\, {\hat R}^{(n)}_{ab}(z,M/Q_0) \;\; .
\end{equation}
The next-to-leading order (NLO) and NNLO contributions to the cumulative cross
section in Eq.~(\ref{inte}) are determined by the 
%perturbative
functions ${\hat R}^{(1)}$ and
${\hat R}^{(2)}$, respectively.
The small-$q_T$ region of the 
%transverse-momentum 
cross section
$d{\hat \sigma}_{ab}/d q_T^2$ is probed by performing the limit $Q_0\ll M$
in Eq.~(\ref{inte}). In this limit, the NLO and NNLO functions 
${\hat R}^{(1)}$ and ${\hat R}^{(2)}$ have the following
behaviour:
\begin{equation}
\label{eqr1s}
{\hat R}^{(1)}_{ab}(z,M/Q_0)=l_0^2 \;{\hat R}_{ab}^{(1; 2)}(z)
+l_0 \;{\hat R}_{ab}^{(1; 1)}(z)
+{\hat R}_{ab}^{(1; 0)}(z)+{\cal O}(Q_0^2/M^2) \;\;,
\end{equation}
%\begin{align}
%\label{eqr2s}
%{\hat R}^{(2)}_{ab}(z,M/Q_0)&=l_0^4\; {\hat R}_{ab}^{(2;4)}(z)
%+l_0^3\; {\hat R}_{ab}^{(2;3)}(z)+l_0^2\; 
%{\hat R}_{ab}^{(2;2)}(z) \nn \\
%&\,+ l_0 \;{\hat R}_{ab}^{(2;1)}(z)
%+ {\hat R}_{ab}^{(2;0)}(z)
%+{\cal O}(Q_0^2/M^2) \;\;,
%\end{align}
\begin{align}
\label{eqr2s}
{\hat R}^{(2)}_{ab}(z,M/Q_0)=l_0^4\; {\hat R}_{ab}^{(2;4)}(z)
+l_0^3\; {\hat R}_{ab}^{(2;3)}(z)+l_0^2\; 
{\hat R}_{ab}^{(2;2)}(z) 
+ l_0 \;{\hat R}_{ab}^{(2;1)}(z)
+ {\hat R}_{ab}^{(2;0)}(z)
+{\cal O}(Q_0^2/M^2) \;,
\end{align}
where $l_0=\ln (M^2/Q_0^2)$. In Eqs.~(\ref{eqr1s}) and (\ref{eqr2s}), the powers
of the large logarithm $l_0$ are produced by the singular (though, integrable)
behaviour of $d{\hat \sigma}_{ab}/d q_T^2$ at small values of $q_T$.
The coefficients ${\hat R}^{(1;m)}$ (with $m \leq 2$) and ${\hat R}^{(2;m)}$
(with $m \leq 4$) of the large logarithms are independent of $Q_0$; these
coefficients depend on the partonic centre--of--mass energy $\hat s$ and,
more precisely, they are functions of the energy fraction $z=M^2/\hat s$.

In this paper we present the computation of the cumulative cross section in
Eq.~(\ref{inte}) up to NNLO. The partonic calculation is performed in analytic
form by neglecting terms of ${\cal O}(Q_0^2/M^2)$ in the limit $Q_0 \ll M$.
Therefore, we determine the coefficient functions ${\hat R}^{(n;m)}(z)$
in Eqs.~(\ref{eqr1s}) and (\ref{eqr2s}). Before presenting the results, we
illustrate how 
%the behaviour in Eqs.~(\ref{eqr1s}) and (\ref{eqr2s}) is 
these 
%coefficient 
functions are
related to the perturbative coefficients of the transverse-momentum resummation
formula for Higgs boson production \cite{Collins:1984kg, Catani:2010pd}.
This relation, which allows us to extract the $q_T$ resummation coefficients
up to ${\cal O}(\as^2)$, also shows that the knowledge of Eq.~(\ref{eqr2s})
is sufficient to fully determine the NNLO {\em rapidity} distribution
of the Higgs boson in the small-$q_T$ region.

The partonic cross section $d{\hat \sigma}_{ab}/d q_T^2$ in Eq.~(\ref{dcross})
can be decomposed in the form 
$d{\hat \sigma}_{ab} = d{\hat \sigma}_{ab}^{({\rm sing})} 
+ d{\hat \sigma}_{ab}^{({\rm reg})}$.
%The partonic cross sections $d{\hat \sigma}_{ab}/d q_T^2$ are computable in
%QCD perturbation theory as power series expansions in $\as(M^2)$.
%We are interested in the perturbative contributions that are large in the 
%small-$q_T$ region $(q_T \ll M)$ and, eventually, singular in the limit 
%$q_T \to 0$.  To this purpose, we can decompose the partonic cross section
%in the form 
%$d{\hat \sigma}_{ab} = d{\hat \sigma}_{ab}^{({\rm sing})} 
%+ d{\hat \sigma}_{ab}^{({\rm reg})}$.
The singular component, $d{\hat \sigma}_{ab}^{({\rm sing})}$,
contains all the contributions that are enhanced
%(or `singular') 
at small $q_T$. These contributions are proportional to
$\delta(q_T^2)$ or to large logarithms\footnote{To be precise,
the logarithms are combined with corresponding `contact' terms, which are
proportional to $\delta(q_T^2)$. These combinations define regularized 
(integrable) `plus
distributions' $\left[\f{1}{q_T^2}\ln^m (M^2/q_T^2)\right]_+$ with respect 
to~$q_T^2$. The cumulative cross section in Eq.~(\ref{inte}) is insensitive to
the precise mathematical definition of these `plus
distributions'.} 
of the type 
$\f{1}{q_T^2}\ln^m (M^2/q_T^2)$.
On the
contrary, the remaining component, $d{\hat \sigma}_{ab}^{({\rm reg})}$,
of the partonic cross section 
is regular 
%(i.e. free of logarithmic terms)
order-by-order in $\as$ as $q_T \to 0$. To be precise, the
integration of $d{\hat \sigma}_{ab}^{({\rm reg})}/dq_T^2$ over the range 
$0 \leq q_T \leq Q_0$ leads to a finite result that, at each
fixed order in $\as$, {\em vanishes} in the limit $Q_0 \to 0$.
Therefore, $d{\hat \sigma}_{ab}^{({\rm reg})}$ only contributes to the terms
of ${\cal O}(Q_0^2/M^2)$ on the right-hand side of 
Eqs.~(\ref{eqr1s}) and (\ref{eqr2s}).
The decomposition of the partonic cross sections can be inserted in
the right-hand side of Eq.~(\ref{dcross}), thus leading to the corresponding
decomposition of the hadronic cross section
$d\sigma/d q_T^2$, namely,
$d\sigma = d\sigma^{({\rm sing})} 
+ d\sigma^{({\rm reg})}$.

We consider the singular component of the Higgs boson $q_T$ cross
section, and we recall its all-order perturbative structure. We directly
refer to the hadronic cross section (rather than the partonic cross sections),
since its structure can be presented 
%in 
by using
a more compact notation.
Moreover, to illustrate the general kinematics of transverse-momentum
resummation, we consider the $q_T$ cross
section  $d\sigma/dy \,dq_T^2$ at fixed value of the rapidity $y$ of the Higgs
boson (the rapidity is defined in the centre--of--mass frame of the two 
colliding hadrons). The transverse-momentum resummation formula for the 
singular component 
of the Higgs boson cross section is 
%\cite{Collins:1984kg, Catani:2000vq, Catani:2010pd}
\cite{Collins:1984kg, Catani:2010pd}
\beeq
\label{qtycrossgg}
&&\!\!\!\!\!\!\f{{d\sigma}^{({\rm sing})}}{dy \,d q_T^2}(y,q_T,M,s) =
\f{M^2}{s} \;
\sigma_H^{(0)}(\as(M^2)) 
%\left[ d\sigma_{gg, \,F}^{(0)} \right]
\int_0^{+\infty} db \;\f{b}{2}  \;J_0(b q_T) \;
  S_g(M,b)\nn \\
&& \;\;\;\; \times \;
\sum_{a_1,a_2} \,
\int_{x_1}^1 \f{dz_1}{z_1} \,\int_{x_2}^1 \f{dz_2}{z_2} 
\; \left[ H^F C_1 C_2 \right]_{gg;\,a_1a_2}
\;f_{a_1/h_1}(x_1/z_1,b_0^2/b^2)
\;f_{a_2/h_2}(x_2/z_2,b_0^2/b^2) \;
\;, 
\eeeq
where the kinematical variables $x_i$ $(i=1,2)$ are 
$x_1= e^{+y} M/{\sqrt s}$ and $x_2= e^{-y} M/{\sqrt s}$. The integration
variable $b$ is the impact parameter, $J_0(b q_T)$ is the $0$th-order 
Bessel function, and $b_0=2e^{-\gamma_E}$
($\gamma_E=0.5772\dots$ is the Euler number) is a numerical coefficient.
The symbol $\left[ H^F C_1 C_2 \right]_{gg;\,a_1a_2}$ denotes
(see Eq.~(44) in Ref.~\cite{Catani:2010pd})
the following 
%pertubative 
function of the longitudinal-momentum fractions
$z_1$ and $z_2$:
%\beeq
%\label{whath}
%\left[ H^{H} C_1 C_2 \right]_{gg;\,a_1a_2}
%&=& H_{g}^{H}(\as(M^2)) \;\left[ \; C_{g \,a_1}(z_1;\as(b_0^2/b^2)) 
%\;\, C_{g \,a_2}(z_2;\as(b_0^2/b^2)) \right.
%\nn \\
%&+& \left.  G_{g \,a_1}(z_1;\as(b_0^2/b^2)) 
%\;\, G_{g \,a_2}(z_2;\as(b_0^2/b^2)) \;
%\right]
%\;\;. 
%\eeeq
\beeq
\label{whath}
\left[ H^{H} C_1 C_2 \right]_{gg;\,a_1a_2}
= H_{g}^{H}(\as(M^2)) \!\!\!\!\!\!&&\left[ \;\; C_{g \,a_1}(z_1;\as(b_0^2/b^2)) 
\;\; C_{g \,a_2}(z_2;\as(b_0^2/b^2)) \right.
\nn \\
&&+ \left.  G_{g \,a_1}(z_1;\as(b_0^2/b^2)) 
\;\, G_{g \,a_2}(z_2;\as(b_0^2/b^2)) \;
\right]
\;\;, 
\eeeq
where $H_{g}^{H}(\as), C_{g \,a}(z;\as)$ and  $G_{g \,a}(z;\as)$ are
perturbative functions of $\as$ (see Eqs.~(\ref{gpert})--(\ref{hpert})).
The other perturbative ingredients of 
Eq.~(\ref{qtycrossgg}) are the function $S_g(M,b)$, which is the Sudakov form
factor of the gluon (see the comments below), and $\sigma_H^{(0)}(\as(M^2))$,
which is the Born level cross section 
in Eq.~(\ref{sig0}).
%for the partonic subprocess $gg\to H$.
%Using the large-$m_{top}$ approximation, the explicit expression of 
%$\sigma_H^{(0)}$ is \cite{Ellis:1975ap}
%\begin{equation}
%\sigma_H^{(0)}(\as)=\f{G_F \;\as^2}{288 \,\pi \,\sqrt{2}} \;\;,
%\end{equation}
%where $G_F$ is the Fermi constant.

The structure of Eq.~(\ref{qtycrossgg}) is well known in the literature
on resummed calculations for the $q_T$ spectrum of the Higgs boson.
However, the functional form of Eq.~(\ref{whath}) is new \cite{Catani:2010pd}.
The customary `\naive' version (i.e. the version extrapolated from the
transverse-momentum resummation formula for the Drell--Yan process)
of Eq.~(\ref{whath}) includes only the perturbative functions 
$H_{g}^{H}$ and $C_{g \,a}$. The presence of an additional term, due to the
function $G_{g \,a}$, has been pointed out in Ref.~\cite{Catani:2010pd}.
Note that the function $G_{g \,a}(z;\as)$ is of ${\cal O}(\as)$ 
(see Eq.~(\ref{gpert})) and, therefore,
it leads to a contribution of ${\cal O}(\as^2)$ in the right-hand side of 
Eq.~(\ref{whath}). This fact implies that the presence of $G_{g \,a}(z;\as)$
cannot be detected through a NLO calculation of the $q_T$ spectrum 
of the Higgs boson. This fact also implies that our NNLO analytic calculation of
the cumulative cross section in Eq.~(\ref{inte}) gives an explicit check of the
presence of $G_{g \,a}(z;\as)$ and of its precise form at ${\cal O}(\as)$
(see Eq.~(\ref{qq2}) and related comments).

The gluon form factor $S_g(M,b)$ of Eq.~(\ref{qtycrossgg}) is a
process-independent quantity \cite{Collins:1984kg, Catani:vd, Catani:2000vq}. 
Its functional
dependence on $M$ and $b$ is controlled by two perturbative functions, which are
usually denoted as $A_g(\as)$ and $B_g(\as)$ (see, e.g.,  
Eqs.~(10)--(12) in Ref.~\cite{Catani:2010pd}). Their corresponding $n$-th
order perturbative coefficients are $A_g^{(n)}$ and $B_g^{(n)}$. The
coefficients $A_g^{(1)}$, $B_g^{(1)}$, $A_g^{(2)}$ \cite{Catani:vd} and
$B_g^{(2)}$ \cite{deFlorian:2001zd} are known: their knowledge fully determines
the perturbative expression of $S_g(M,b)$ up to ${\cal O}(\as^2)$.

%%%%%%%%%%%%%%%%%%
\setcounter{footnote}{1}
%%%%%%%%%%%%%%%%%%

The quantity $\left[ H^F C_1 C_2 \right]$ in Eq.~(\ref{qtycrossgg}) depends on
the three perturbative functions
$H_{g}^{H}, C_{g \,a}$ and  $G_{g \,a}$. By inspection of the right-hand side of 
Eq.~(\ref{whath}), we notice that the scale of $\as$ is not set to a unique
value. We have $\as(M^2)$ in the case of the function $H_{g}^{H}(\as)$,
and $\as(b_0^2/b^2)$ in the case of the functions $C_{g \,a}(\as)$ and  
$G_{g \,a}(\as)$. The presence of these two different arguments of $\as$ is
related to the physical origin \cite{Catani:2000vq, Catani:2010pd} 
of the corresponding perturbative functions. Roughly speaking,
$H_{g}^{H}(\as(M^2))$ embodies contributions due to the {\em hard}-momentum
region\footnote{This is the region where the size of the momenta of the virtual
loops is of the order of $M$.}
of the {\em virtual} corrections to the lowest-order subprocess $gg \to H$.
The functions  $C_{g \,a}$ and  $G_{g \,a}$ instead refer to the inclusive
subprocess $g\,a \to H + X$: roughly speaking,
$C_{g \,a}(\as(b_0^2/b^2))$ and  $G_{g \,a}(\as(b_0^2/b^2))$ originate
from the kinematical region where the momenta of the partons in the final-state
system $X$ are (almost) {\em collinear} to the momentum of the initial-state
parton $a$. Owing to this physical picture,
%intuitive origin, 
the quantity 
$\left[ H^F C_1 C_2 \right]$ can be regarded as a {\em hard-collinear} 
partonic function. Note that the function $H_{g}^{H}(\as)$ is process dependent,
since it is directly related to the production mechanism of the SM Higgs boson.
On the contrary, the partonic functions $C_{g \,a}$ and  $G_{g \,a}$ are process
independent, as a consequence of the universality features of QCD collinear
radiation.

We recall that
%However, 
the functions 
$H_{g}^{H}(\as), C_{g \,a}(\as), G_{g \,a}(\as)$ and the perturbative function
$B_g(\as)$ of the gluon form factor are not {\em separately} computable in an
unambiguous way. Indeed, these four functions are correlated (constrained) by a
renormalization-group symmetry \cite{Catani:2000vq} that is related to the 
$b$-space factorization structure of Eqs.~(\ref{qtycrossgg}) and (\ref{whath}). 
The unambiguous definition of these four functions thus requires the
specification\footnote{The reader who is not interested in issues related to the
specification of a resummation scheme can simply assume that
$H_{g}^{H}(\as) \equiv 1$ throughout this paper.}
of a {\em resummation scheme} \cite{Catani:2000vq}.
Note, however, that considering the perturbative 
expansion\footnote{The resummation-scheme dependence also cancels
by consistently expanding Eq.~(\ref{qtycrossgg}) in terms of classes of
resummed (leading, next-to-leading and so forth) logarithmic contributions
\cite{Bozzi:2005wk}.}
of Eq.~(\ref{qtycrossgg}) (i.e., the perturbative expansion of the singular
component of the $q_T$ cross section), the resummation-scheme dependence exactly
cancels order-by-order in $\as$. 

The perturbative expansion of the three functions on the right-hand side of
Eq.~(\ref{whath}) is defined as follows:
\begin{equation}
\label{gpert}
G_{g\,a}(z; \as)= \frac{\as}{\pi} \;G_{g\,a}^{(1)}(z)+
\sum_{n=2}^\infty \left(\frac{\as}{\pi}\right)^n G_{g\,a}^{(n)}(z) \;\; ,
\end{equation}
\begin{equation}
\label{cpert}
C_{g\,a}(z; \as)=\delta_{g\,a} \;\delta(1-z)+
\sum_{n=1}^{\infty}\left(\frac{\as}{\pi}\right)^n C_{g\,a}^{(n)}(z) \;\; ,
\end{equation}
\begin{equation}
\label{hpert}
H_g^H(\as) = 1+\sum_{n=1}^\infty \left(\frac{\as}{\pi}\right)^n H_g^{H(n)}\, .
\end{equation}
Since the partonic functions $G_{g\,a}$ and $C_{g\,a}$ are process independent,
they fulfil the following relations:
\beq
\label{fdep}
G_{g \,q_f}(z;\as) = G_{g \,{\bar q}_{f'}}(z;\as) \equiv G_{g \,q}(z;\as) \;\;,
\quad \;\;
C_{g \,q_f}(z;\as) = C_{g \,{\bar q}_{f'}}(z;\as) \equiv C_{g \,q}(z;\as) \;\;,
\eeq
which are a consequence of charge conjugation invariance and 
flavour symmetry of QCD. The dependence of $G_{g\,a}$ ($C_{g\,a}$) on the parton
label $a$ is thus fully specified by $G_{g\,g}$ and $G_{g\,q}$ ($C_{g\,g}$ and
$C_{g\,q}$). The first-order coefficient functions $G_{g\,g}^{(1)}(z)$ 
and $G_{g\,q}^{(1)}(z)$
(they are independent of the resummation scheme) are 
known \cite{Catani:2010pd}:
\beq
\label{gone}
G_{g\,g}^{(1)}(z) = C_A \;\f{1-z}{z} \;\;, 
\quad \;\;\;\;
G_{g\,q}^{(1)}(z) = C_F \;\f{1-z}{z} \;\;.
\eeq
%The knowledge of the first-order
%coefficients in Eq.~(\ref{gone}) has a non-trivial role in the context of 
%the NNLO calculation presented in this paper.
The first-order coefficient function $C_{g\,q}^{(1)}(z)$ is also independent on
the resummation scheme; its expression is 
\cite{Kauffman:1991cx}
%\cite{Kauffman:1991cx, deFlorian:2001zd}:
\beq
\label{conegq}
C_{g\,q}^{(1)}(z) = \f{1}{2} \;C_F \;z \;\;. 
\eeq
%%%%-----------
Using the large-$m_{top}$ approximation, the first-order coefficients 
$H_g^{H(1)}$ and $C_{g\,g}^{(1)}(z)$ fulfil the following relation
\cite{deFlorian:2001zd}:
\beq
\label{conegg}
C_{g\,g}^{(1)}(z) + \f{1}{2} \,H_g^{H(1)} \,\delta(1-z) 
=  \f{(5 + \pi^2) C_A - 3 C_F}{4} \;\,\delta(1-z) \;\;.
\eeq
The separate determination of $C_{g\,g}^{(1)}(z)$ and $H_g^{H(1)}$ requires the
specification of a resummation scheme. For instance, considering the 
resummation scheme in which the SM Higgs boson coefficient $H_g^{H(1)}$
vanishes, the right-hand side of Eq.~(\ref{conegg}) gives the value of 
$C_{g\,g}^{(1)}(z)$ \cite{Kauffman:1991cx}, and the corresponding value 
of the gluon form factor coefficient $B_g^{(2)}$ is explicitly reported
in Eq.~(128) of the second paper in Ref.~\cite{deFlorian:2001zd}.
%The separate determination of $C_{g\,g}^{(1)}(z)$ and $H_g^{H(1)}$ requires the
%specification of a resummation scheme. For instance, considering the 
%resummation scheme\footnote{Using this resummation scheme, the corresponding
%value of the gluon form factor coefficient $B_g^{(2)}$ is explicitly reported
%in Eq.~(128) of the second paper in Ref.~\cite{deFlorian:2001zd}.}
%in which the SM Higgs boson coefficient $H_g^{H(1)}$
%vanishes (in the large-$m_{top}$ approximation), we have\footnote{****
%dependence on colour factors TO BE CHECKED ****}
%\cite{Kauffman:1991cx, deFlorian:2001zd}
%\beq
%\label{conegg}
%C_{g\,g}^{(1)}(z) =  \f{(5 + \pi^2) C_A - 3 C_F}{4} \;\,\delta(1-z) \;\;,
%\quad \;\;\; H_g^{H(1)}= 0 \;\;.
%\eeq
%%%%%%--------
The computation of the second-order coefficients $C_{g\,q}^{(2)}, 
C_{g\,g}^{(2)}$ and $H_g^{H(2)}$ is the 
%purpose 
aim of the calculation
described in this 
%the present 
paper.

For later purposes, we also define the following hard-collinear coefficient
function:
\begin{equation}
\label{HCCGG}
{\cal H}^H_{gg\ito ab}(z;\as) \equiv H_g^H(\as) \!\int_0^1\!dz_1 \int_0^1\!dz_2
\,\delta(z - z_1z_2)
\Big[ C_{g \,a}(z_1;\as) C_{g \,b}(z_2;\as)
+ G_{g \,a}(z_1;\as) G_{g \,b}(z_2;\as)\Big]\, ,
\end{equation}
%\begin{equation}
%\label{HCCGG}
%{\cal H}^H_{gg\ito ab}(z;\as) \equiv H_g^H(\as) \int_0^1 dz_1 \,dz_2
%\,\delta(z - z_1z_2)
%\Big[ C_{g \,a}(z_1;\as) C_{g \,b}(z_2;\as)
%+ G_{g \,a}(z_1;\as) G_{g \,b}(z_2;\as)\Big]\, ,
%\end{equation}
which is directly related to the coefficient function in Eq.~(\ref{whath}).
There are only two differences between Eqs.~(\ref{whath}) and (\ref{HCCGG}).
The first difference is due to the fact that the function ${\cal H}^H$ depends on
the energy fraction $z$, since the right-hand side of Eq.~(\ref{HCCGG})
involves a convolution integral over the momentum fractions $z_1$ and $z_2$.
This convolution kinematically arises by considering the integration of 
Eq.~(\ref{qtycrossgg}) over the rapidity $y$ of the Higgs boson.
The second difference regards the scale of $\as$: in the functions 
$H^H(\as), C(\as)$ and $G(\as)$ on the right-hand side of Eq.~(\ref{HCCGG}),
the argument of $\as$ is set to the same value (this common scale is not
explicitly denoted in Eq.~(\ref{HCCGG})). Owing to this feature, the
process-dependent function ${\cal H}^H_{gg\ito ab}$ is unambiguously defined
(i.e., it is independent of the specification of the resummation scheme) 
\cite{Catani:2000vq}. The perturbative expansion of the function
${\cal H}^H$ directly follows from Eqs.~(\ref{gpert})--(\ref{hpert}). We have:
\begin{equation}
\label{chpert}
{\cal H}^H_{gg\ito ab}(z;\as) =\delta_{g\,a} \,\delta_{g\,b}\;\delta(1-z)+
\sum_{n=1}^{\infty}\left(\frac{\as}{\pi}\right)^n 
{\cal H}^{H(n)}_{gg\ito ab}(z) \;\; ,
\end{equation}
where the first-order and second-order contributions are
\begin{equation}
\label{H1}
{\cal H}^{H(1)}_{gg\ito ab}(z)=\delta_{g\,a} \,\delta_{g\,b} \,\delta(1-z) 
\,H^{H(1)}_g+
\delta_{g\,a} \,C^{(1)}_{g\,b}(z)+\delta_{g\,b} \,C^{(1)}_{g\,a}(z) \;\; ,
\end{equation}
\begin{align}
\label{H2}
{\cal H}^{H(2)}_{gg\ito ab}(z)&=\delta_{g\,a} \,\delta_{g\,b} \,\delta(1-z)
\,H^{H(2)}_g
+\delta_{g\,a} \,C^{(2)}_{g\,b}(z)+\delta_{g\,b} \,C^{(2)}_{g\,a}(z)
+H^{H(1)}_g\left(\delta_{g\,a} \,C^{(1)}_{g\,b}(z)
+\delta_{g\,b} \,C^{(1)}_{g\,a}(z)\right)\nn\\
&
+\left(C^{(1)}_{g\,a}\otimes C^{(1)}_{g\,b}\right)(z)+
\left(G^{(1)}_{g\,a}\otimes G^{(1)}_{g\,b}\right)(z)\;\; .
\end{align}
In Eq.~(\ref{H2}) and in the following, the symbol $\otimes$ denotes the 
convolution integral
(i.e., we define 
$(g \otimes h) (z) \equiv \int_0^1 dz_1 \int_0^1 dz_2
\,\delta(z - z_1z_2) \;g(z_1) \;h(z_2)$).

%%%%%%%%%%%%%%%%%%
\setcounter{footnote}{2}
%%%%%%%%%%%%%%%%%%

After our 
illustration of the all-order resummation formula in
Eq.~(\ref{qtycrossgg}), we can return to its relation with the perturbative 
expression of the cumulative partonic cross section in Eq.~(\ref{inte}).
Using the Altarelli--Parisi evolution equations, the parton densities
$f_{a/h}(x,b_0^2/b^2)$ on the right-hand side of Eq.~(\ref{qtycrossgg})
can be expressed in terms of the corresponding parton densities
$f_{a/h}(x,M^2)$ at the factorization (evolution) scale $\mu_F=M$.
Having done that, all the remaining factors in Eq.~(\ref{qtycrossgg})
are the partonic contributions that determine the small-$q_T$ singular 
component of the Higgs boson partonic cross section 
$d{\hat \sigma}_{ab}/d q_T^2$ in Eqs.~(\ref{dcross}) and (\ref{inte}).
At fixed values of the impact parameter $b$, all these partonic
contributions can be 
%perturbatively 
expanded in powers of 
$\as(M^2)$, thus leading to perturbative coefficients that depend on 
powers of $\ln (b^2M^2)$. The dependence on $\ln (b^2M^2)$ is produced
by the gluon form factor $S_g(M,b)$, by the Altarelli--Parisi evolution 
equations and by the QCD coupling\footnote{The coupling
$\as(b_0^2/b^2)$ can be expressed in terms of $\as(M^2)$ and 
$\ln (b^2M^2/b_0^2)$ by using the renormalization group equation for the
perturbative $\mu^2$-evolution of the 
%QCD 
running coupling $\as(\mu^2)$.} 
$\as(b_0^2/b^2)$. The powers of $\ln (b^2M^2)$ can then be transformed into
logarithms, $\ln (M^2/q_T^2)$, in $q_T$-space by explicitly performing the
Bessel transformation (i.e. the integration over $b$) in Eq.~(\ref{qtycrossgg}).
This procedure, which involves manipulations that are standard in the context of
transverse-momentum resummation (technical details can be found, for instance,
in Ref.~\cite{Bozzi:2005wk}), yields the explicit perturbative
expression of the singular component of 
$d{\hat \sigma}_{ab}/d{\hat y} d q_T^2\,$.
The integration of this expression over ${\hat y}$ (${\hat y}$ is the rapidity
of the Higgs boson in the 
%partonic 
centre--of--mass frame 
of the two colliding partons)
and $q_T$ finally gives the cumulative partonic cross section of 
Eqs.~(\ref{inte}) and (\ref{eqnR}) in the limit $Q_0 \ll M$
(i.e., modulo the contributions of ${\cal O}(Q_0^2/M^2)$ in 
Eqs.~(\ref{eqr1s}) and (\ref{eqr2s})).
The perturbative functions ${\hat R}^{(n)}_{ab}(z,M/Q_0)$ of Eq.~(\ref{eqnR})
have a dependence on $l_0=\ln(M^2/Q_0^2)$ that is explicitly determined by
the resummation formula (\ref{qtycrossgg}), whereas the dependence on $z$ is
given in terms of the $q_T$ resummation coefficients (those of the gluon form
factor and 
%those 
in Eqs.~(\ref{gpert})--(\ref{hpert})).

The NLO and NNLO functions 
${\hat R}^{(1)}_{ab}$ and ${\hat R}^{(2)}_{ab}$ have the following
expressions:
\begin{equation}
\label{eqr1}
{\hat R}^{(1)}_{ab}(z,M/Q_0)=l_0^2 \;\Sigma_{gg \ito ab}^{H(1;2)}(z)
+l_0\,\Sigma_{gg \ito ab}^{H(1;1)}(z)+{\cal H}_{gg\ito ab}^{H(1)}(z)
+{\cal O}(Q_0^2/M^2) \;\;,
\end{equation}
\begin{align}
\label{eqr2}
{\hat R}^{(2)}_{ab}(z,M/Q_0)&=l_0^4 \;\Sigma_{gg\ito ab}^{H(2;4)}(z)
+l_0^3\; \Sigma_{gg\ito ab}^{H(2;3)}(z)+l_0^2\; \Sigma_{gg\ito ab}^{H(2;2)}(z)
+l_0\left(\Sigma_{gg\ito ab}^{H(2;1)}(z)-16 \zeta_3 \Sigma_{gg\ito ab}^{H(2;4)}(z)\right)
\nn\\
&
+\left({\cal H}_{gg\ito ab}^{H(2)}(z)
-4\zeta_3\, \Sigma_{gg\ito ab}^{H(2;3)}(z)\right)
+{\cal O}(Q_0^2/M^2)\, ,
\end{align}
which are consistent with the behaviour in Eqs.~(\ref{eqr1s}) and (\ref{eqr2s}).
In Eqs.~(\ref{eqr1}) and (\ref{eqr2}) we use the same notation as in 
Ref.~\cite{Bozzi:2005wk}. The coefficient functions 
$\Sigma_{gg\ito ab}^{H(n;m)}(z)$
%%and ${\cal H}_{gg\ito ab}^{H(n)}(z)$
depend on the $q_T$ resummation coefficients: the explicit expressions are given
%%in Eqs.~(63)--(70) of Ref.~\cite{Bozzi:2005wk}
in Eqs.~(63),(64),(66)--(69) of Ref.~\cite{Bozzi:2005wk}
(we have to set $\mu_R=\mu_F=Q=M$, where $\mu_R, \mu_F$ and $Q$
are the auxiliary scales of Ref.~\cite{Bozzi:2005wk})
and are not reported here.
The coefficients
${\cal H}_{gg\ito ab}^{H(1)}$ and ${\cal H}_{gg\ito ab}^{H(2)}$ are exactly 
those in Eqs.~(\ref{H1}) and (\ref{H2}).
The first-order terms 
$\Sigma_{gg\ito ab}^{H(1;2)}$ and $\Sigma_{gg\ito ab}^{H(1;1)}$ depend
on the gluon form factor $S_g(M,b)$. The second-order terms
$\Sigma_{gg\ito ab}^{H(2;m)}$ 
%(with $m=1,2,3,4$) 
depend on 
${\cal H}_{gg\ito ab}^{H(1)}$ and on the gluon form factor $S_g(M,b)$
up to ${\cal O}(\as^2)$. The numerical coefficient 
$\zeta_3 \simeq 1.202\dots$ ($\zeta_k$ is the Riemann $\zeta$-function) 
on the right-hand side of Eq.~(\ref{eqr2}) originates from the Bessel
transformations 
%from $b$-space to $q_T$-space 
(see, e.g., Eqs.~(B.18) and (B.30)
%i.e. Eqs.~(129) and (141) in the arXiv version 
in Appendix~B of Ref.~\cite{Bozzi:2005wk}).

The relations (\ref{eqr1}) and (\ref{eqr2})
can be exploited in two different ways. From the knowledge of the perturbative
coefficients of the resummation formulae (\ref{qtycrossgg}) and (\ref{whath}),
we can compute $\Sigma^{H(n;m)}$ and ${\cal H}^{H(n)}$ and then,
we can obtain a perturbative prediction for the cumulative partonic cross
section up to NNLO. Alternatively, from the explicit NNLO perturbative 
computation of the cumulative partonic cross section, we can extract 
$\Sigma^{H(n;m)}$ and ${\cal H}^{H(n)}$ and then, we can determine the 
$q_T$ resummation coefficients up to ${\cal O}(\as^2)$.

\newpage

Our NNLO computation of the cumulative partonic cross section is described in the
final part of this paper. We obtain the following results.
The explicit result of the NLO function ${\hat R}^{(1)}_{ab}(z)$ confirms the
expressions of $\Sigma_{gg\ito ab}^{H(1;2)}(z)$,  
$\Sigma_{gg\ito ab}^{H(1;1)}(z)$ and ${\cal H}^{H(1)}_{gg\ito ab}(z)$, as
predicted by the $q_T$ resummation coefficients at ${\cal O}(\as)$.
At NNLO, the present knowledge \cite{Catani:vd, deFlorian:2001zd} of the 
$q_T$ resummation coefficients at ${\cal O}(\as^2)$ predicts the expressions of
the terms $\Sigma_{gg\ito ab}^{H(2;m)}(z)$, with $m=1,2,3,4$. Our result for the
NNLO function ${\hat R}^{(2)}_{ab}(z)$ confirms this prediction, and it allows us
to extract the explicit expression of the second-order coefficient function 
${\cal H}^{H(2)}_{gg\ito ab}(z)$.

%\newpage

We obtain:
\begin{equation}
\label{h2qq}
{\cal H}^{H(2)}_{gg\ito qq}(z)
=- \,C^2_F\left[\f{2(1-z)}{z}+\f{(2+z)^2}{4z}\ln z\right]\, ,
\end{equation}
\begin{align}
\label{h2gq}
{\cal H}^{H(2)}_{gg\ito gq}(z)&=C_F^2\Bigg(
\f{1}{48}\left(2-z\right) \ln^3 z
-\f{1}{32}\left(3z+4\right) \ln^2 z
+\f{5}{16}\left(z-3\right) \ln z\nn\\
&~~~~~~+\f{1}{12}\left(\f{1}{z}+\f{z}{2}-1\right)\ln^3 (1-z)
+\f{1}{16}\left(z+\f{6}{z}-6\right) \ln^2(1-z)\nn\\
&~~~~~~+\left(\f{5z}{8}+\f{2}{z}-2\right) \ln (1-z)+\f{5}{8}-\f{13}{16}z\Bigg)\nn\\
&+C_F\, n_F \Bigg(\f{1}{24z}\left(1+(1-z)^2\right)\ln^2(1-z)
+\f{1}{18}\left(z+\f{5}{z}-5\right)\ln(1-z)\nn\\
&~~~~~~-\f{14}{27}+\f{14}{27z}+\f{13}{108}z\Bigg)\nn\\
&+C_FC_A\Bigg(-\f{(1+(1+z)^2)}{2z}{\rm Li}_3\left(\f{1}{1+z}\right)
+\left(\f{1}{2}-\f{5}{2z}-\f{5}{4}z\right){\rm Li}_3(z)\nn\\
&-\f{3}{4z}\left(1+(1+z)^2\right){\rm Li}_3(-z)
+\left(2-\f{11}{6z}-\f{z}{2}+\f{z^2}{3}+\left(-\f{1}{2}+\f{3}{2z}+\f{3z}{4}\right)\ln z\right){\rm Li}_2(z)\nn\\
&+\left(\f{z}{4}+\f{(1+(1+z)^2)}{4z}\ln(z)\right){\rm Li}_2(-z)\nn\\
&+\f{\left(1+(1+z)^2\right)}{12z}\ln^3(1+z)
%+\f{1}{24z}\Big(\left(1+(1+z)^2\right)\left(3\ln^2 z-\pi^2\right)-6z^2 \ln z\Big)\ln(1+z)\nn\\
-\f{1}{24z}\Big(\left(1+(1+z)^2\right)\left(3\ln^2 z+\pi^2\right)-6z^2 \ln z\Big)\ln(1+z)\nn\\
&-\f{\left(1+(1-z)^2)\right)}{24z}\ln^3(1-z)
+\f{1}{48z}\left(6(1+(1-z)^2)\ln z-5z^2-22(1-z)\right)\ln^2(1-z)\nn\\
&+\f{1}{72z}\left(-152 + 152 z - 43 z^2 + 6(-22 + 24 z - 9 z^2 + 4 z^3)\ln z
+ 9(1+(1-z)^2)\ln^2 z\right)\ln(1-z)\nn\\
&-\f{1}{12}\left(1+\f{z}{2}\right)\ln^3 z+\f{1}{48}\left(36+9z+8z^2\right)\ln^2 z+\left(-\f{107}{24}-\f{1}{z}+\f{z}{12}-\f{11}{9}z^2\right)\ln z\nn\\
&+\f{1}{z}\left(4\zeta_3-\f{503}{54}+\f{11}{36}\pi^2\right)
+\f{1007}{108}-\f{\pi^2}{3}-\f{5}{2}\zeta_3
+z\left(\f{\pi^2}{3}+2\zeta_3-\f{133}{108}\right)
+z^2\left(\f{38}{27}-\f{\pi^2}{18}\right)\Bigg)\, ,
\end{align}
\begin{align}
\label{h2gg}
% Mismatch in colour structure in delta[1-z] term corrected
{\cal H}^{H(2)}_{gg\ito gg}(z)&=
\Bigg(\left(-\f{101}{27}+\f{7}{2}
\,\zeta_3\right)C_A^2+\f{14}{27}C_A\, n_F\Bigg)
%{\cal D}_0(z)\nn\\
\left(\f{1}{1-z} \right)_+ \nn\\
&+\Bigg(C_A^2\left(
\f{3187}{288}+\f{7}{8}L_t+\f{157}{72}\pi^2+\f{13}{144}\pi^4-\f{55}{18}\zeta_3\right)+C_A\, C_F\left(-\f{145}{24}-\f{11}{8}L_t-\f{3}{4}\pi^2\right)\nn\\
&~~~+\f{9}{4}C_F^2
-\f{5}{96}C_A-\f{1}{12}C_F-C_A\, n_F\left(\f{287}{144}+\f{5}{36}\pi^2+\f{4}{9}\zeta_3\right)\nn\\
&~~~+C_F\, n_F\left(-\f{41}{24}+\f{1}{2}L_t+\zeta_3\right)\Bigg)\delta(1-z)\nn\\
&~~~+C_A^2\Bigg(\f{(1+z+z^2)^2}{z(1+z)}\left(2{\rm Li}_3\left(\f{z}{1+z}\right)-{\rm Li}_3(-z)\right)
+\f{2-17z-22z^2-10z^3-12z^4}{2z(1+z)}\zeta_3\nn\\
&~~~-\f{5-z+5z^2+z^3-5z^4+z^5}{z(1-z)(1+z)} \left({\rm Li}_3(z)-\zeta_3\right)
+{\rm Li}_2(z)\,\f{\ln(z)}{1-z}\, \f{3-z+3z^2+z^3-3z^4+z^5}{z(1+z)}\nn\\
&~~~+\f{(1+z+z^2)^2}{z(1+z)}
\left(\ln(z){\rm Li}_2(-z)-\f{1}{3}\ln^3(1+z)+\zeta_2\, \ln(1+z)
\right)\nn\\
&~~~+\f{1-z}{3z} (11-z+11z^2){\rm Li}_2(1-z)+
\f{1}{12} z \ln(1-z)-
\f{1}{6} \f{\ln^3(z)}{1-z} \f{(1+z-z^2)^2}{1+z}\nn\\
&~~~+\ln^2(z) \left(\f{(1-z+z^2)^2}{2z(1-z)}\ln(1-z)-\f{(1+z+z^2)^2}{2z(1+z)}\ln(1+z)+
\f{25-11z+44z^2}{24}\right)\nn\\
&~~~+\ln(z) \left(\f{(1+z+z^2)^2}{z(1+z)}\ln^2(1+z)+\f{(1-z+z^2)^2}{2z(1-z)} \ln^2(1-z)\right. \nn\\
&~~~-\left.\f{72+773z+149 z^2+536 z^3}{72z}\right)
+\f{517}{27}-\f{449}{27z}-\f{380z}{27}+\f{835 z^2}{54}
\Bigg)\nn\\
&+C_A\, n_F\Bigg(\f{1+z}{12}\ln^2(z)+\f{1}{36}(13+10 z)\ln(z)-\f{z}{12}\ln(1-z)-\f{83}{54}+\f{121}{108z}+\f{55}{54} z-\f{139}{108}z^2\Bigg)\nn\\
&+C_F\,
n_F\Bigg(\f{1+z}{12}\ln^3(z)+\f{1}{8}(3+z)\ln^2(z)+\f{3}{2}(1+z)\ln(z)-\f{1-z}{6z}(1-23z+z^2)\Bigg)\,
,
\end{align}
where $L_t=\ln(M^2/m_{top}^2)$ ($m_{top}$ is the pole mass of the top quark)
and ${\rm Li}_k(z)$ $(k=2,3)$ are the usual
polylogarithm functions,
\beq
{\rm Li}_2(z)= - \int_0^z \f{dt}{t} \;\ln(1-t) \;\;,
\quad \quad {\rm Li}_3(z)=  \int_0^1 \f{dt}{t} \;\ln(t)\;\ln(1-zt) \;\;.
\eeq

We comment on the Higgs boson results in Eqs.~(\ref{h2qq})--(\ref{h2gg}) 
and on the ensuing determination
of the second-order 
%resummation 
coefficients 
$C_{g\,q}^{(2)}, C_{g\,g}^{(2)}$ and $H_g^{H(2)}$ in Eqs.~(\ref{cpert})
and  (\ref{hpert}).

Considering the dependence on 
%the flavour of 
the parton indices $a$ and $b$,
the hard-collinear function
${\cal H}^{H}_{gg\ito ab}$ is `flavour blind', namely, it fulfils the relations
\begin{equation}
{\cal H}^{H}_{gg\ito q_fq_{f'}}={\cal H}^{H}_{gg\ito q_f{\bar q}_{f'}}
={\cal H}^{H}_{gg\ito {\bar q}_f q_{f'}}=
{\cal H}^{H}_{gg\ito {\bar q}_{f}{\bar q}_{f'}}
\equiv {\cal H}^{H}_{gg\ito qq} \;\;,
\end{equation}
\begin{equation}
{\cal H}^{H}_{gg\ito q_f g}={\cal H}^{H}_{gg\ito g q_f}=
{\cal H}^{H}_{gg\ito g{\bar q}_{f'}}
={\cal H}^{H}_{gg\ito {\bar q}_{f'}g}
\equiv {\cal H}^{H}_{gg\ito gq} \;\;.
\end{equation}
These symmetry relations follows from Eq.~(\ref{fdep})
and from the convolution integral in Eq.~(\ref{HCCGG})
(the convolution integral implies that ${\cal H}^{H}_{gg\ito ab}$
is symmetric with respect to the exchange $a \leftrightarrow b$).
Therefore, the parton matrix ${\cal H}^{H(2)}_{gg\ito ab}$ is completely specified
by the three entries in Eqs.~(\ref{h2qq})--(\ref{h2gg}): the quark--quark
function ${\cal H}^{H(2)}_{gg\ito qq}$, the gluon--quark
function ${\cal H}^{H(2)}_{gg\ito gq}$ and  the gluon--gluon
function ${\cal H}^{H(2)}_{gg\ito gg}$.

Using Eq.~(\ref{H2}), in the quark--quark channel we have
\begin{equation}
\label{qq2}
{\cal H}^{H(2)}_{gg\ito qq}(z)=
\left(C^{(1)}_{g\,q}\otimes C^{(1)}_{g\,q}\right)(z)+
\left(G^{(1)}_{g\,q}\otimes G^{(1)}_{g\,q}\right)(z) \;\; .
\end{equation}
We see that the second-order coefficient function 
${\cal H}^{H(2)}_{gg\ito qq}(z)$ is fully determined by the $q_T$
resummation coefficients at ${\cal O}(\as)$. Using the values of 
$G^{(1)}_{g\,q}$ and $C^{(1)}_{g\,q}$ in Eqs.~(\ref{gone}) and (\ref{conegq}),
the expression on the right-hand side of Eq.~(\ref{qq2}) is in complete 
agreement with the result in Eq.~(\ref{h2qq}).
Therefore, our explicit computation of the NNLO partonic function
${\hat R}^{(2)}_{qq}$ represents a consistency check of the factorization 
formula
(\ref{whath}) and of the value of $G^{(1)}_{g\,q}$, which were derived in the
process-independent  study of Ref.~\cite{Catani:2010pd}.

Considering the gluon--quark channel, Eq.~(\ref{H2}) can be recast in the
following form:
\begin{equation}
\label{gq2}
C^{(2)}_{g\,q}(z) + \f{1}{2} \,H^{H(1)}_g \,C^{(1)}_{g\,q}(z) =
{\cal H}^{H(2)}_{gg\ito gq}(z) 
- \f{1}{2} \left( {\cal H}^{H(1)}_{gg\ito gg} \otimes C^{(1)}_{g\,q}
\right)(z)
- \left(G^{(1)}_{g\,g}\otimes G^{(1)}_{g\,q}\right)(z)\;\; ,
\end{equation}
where we have used
${\cal H}^{H(1)}_{gg\ito gg}(z) = H^{H(1)}_g \,\delta(1-z) 
+ 2 \,C^{(1)}_{g\,g}(z)$ (see Eq.~(\ref{H1})).
The relation (\ref{gq2}) can be used to determine $C^{(2)}_{g\,q}(z)$ from the
knowledge of ${\cal H}^{H(2)}_{gg\ito gq}$ and of the 
$q_T$ resummation coefficients at ${\cal O}(\as)$
(in particular, the values of $G^{(1)}_{g\,g}$ and $G^{(1)}_{g\,q}$ affect the
determination of $C^{(2)}_{g\,q}$).
Inserting the first-order results of Eqs.~(\ref{gone})--(\ref{conegg})
in Eq.~(\ref{gq2}), we explicitly have:
\begin{equation}
\label{gq22}
C^{(2)}_{g\,q}(z) + \f{1}{4} \,H^{H(1)}_g \,C_F \,z =
{\cal H}^{H(2)}_{gg\ito gq}(z) + C_F^2 \,\f{3}{8} \,z
+ C_F C_A \f{1}{z} \left[ (1+z) \ln z + 2(1-z) - \f{5+\pi^2}{8} \,z^2
\right] \;,
\end{equation}
where ${\cal H}^{H(2)}_{gg\ito gq}$ is given in Eq.~(\ref{h2gq}).
Note that the right-hand side of Eq.~(\ref{gq2}) (or Eq.~(\ref{gq22}))
is resummation-scheme independent. Analogously to Eq.~(\ref{conegg}),
the dependence of $C^{(2)}_{g\,q}$ on the resummation scheme is thus parametrized
by the first-order coefficient $H^{H(1)}_g$ on the left-hand side of 
Eq.~(\ref{gq22}).

The process-independent coefficient function  $C^{(2)}_{g\,g}(z)$ is obtained
analogously to $C^{(2)}_{g\,q}(z)$. Considering the gluon--gluon channel,
Eq.~(\ref{H2}) gives:
\begin{align}
\label{gg2}
2 \,C^{(2)}_{g\,g}(z) &+ \delta(1-z) \left[ H^{H(2)}_g - \f{3}{4} 
\left(H^{H(1)}_g\right)^2 \right] +
\f{1}{2} \,H^{H(1)}_g \,{\cal H}^{H(1)}_{gg\ito gg}(z) \nn \\
&={\cal H}^{H(2)}_{gg\ito gg}(z) 
- \f{1}{4} \left( {\cal H}^{H(1)}_{gg\ito gg} \otimes 
{\cal H}^{H(1)}_{gg\ito gg}\right)(z)
- \left(G^{(1)}_{g\,g}\otimes G^{(1)}_{g\,g}\right)(z)\;\; ,
\end{align}
where the right-hand side is expressed in terms of resummation-scheme 
independent functions. Inserting Eqs.~(\ref{gone})--(\ref{conegg}) in 
Eq.~(\ref{gg2}), we explicitly obtain:
\begin{align}
\label{gg22}
2 \,C^{(2)}_{g\,g}(z) &+ \delta(1-z) \left[ H^{H(2)}_g - \f{3}{4} 
\left(H^{H(1)}_g\right)^2 + \f{(5+\pi^2) C_A - 3 C_F}{4} \,H^{H(1)}_g \right] +
 \nn \\
&={\cal H}^{H(2)}_{gg\ito gg}(z) 
-  \delta(1-z) \left[ \f{(5+\pi^2) C_A - 3 C_F}{4} \right]^2
+ C_A^2 \f{1}{z} \left[ (1+z) \ln z + 2(1-z) \right] \;\; ,
\end{align}
where ${\cal H}^{H(2)}_{gg\ito gg}$ is given in Eq.~(\ref{h2gg}).
We observe that $C^{(2)}_{g\,g}(z)$ includes a resummation-scheme dependent part
that is simply proportional to $\delta(1-z)$. This part depends on 
$H^{H(1)}_g$ and $H^{H(2)}_g$. We also recall \cite{Catani:2000vq} that the 
resummation-scheme invariance relates $C^{(2)}_{g\,g}$, $H^{H(2)}_g$ and the
third-order coefficient $B^{(3)}_g$ of the gluon form factor.

We conclude the paper by briefly describing the method that we have used to
perform the NNLO analytic computation of the Higgs boson cross section in
Eq.~(\ref{inte}). The NNLO partonic calculation has to be carried out by using
dimensional regularization to evaluate the QCD scattering amplitudes and their
integration over the partonic phase space.
%Using the 
In the framework of the 
large-$m_{top}$ approximation, the relevant partonic subprocesses are:
the gluon fusion subprocess $gg \to H$ up to the two-loop level, the
single-emission subprocesses $ab \to H+c$ up to the one-loop level,
and the double-emission subprocesses $ab \to H+c_1+c_2$ at the tree level.
The corresponding scattering amplitudes are known and have been used
in the analytic calculations of two relevant Higgs boson observables:
the NNLO total cross section ${\hat \sigma}_{ab}^{({\rm tot})}$ 
\cite{Harlander:2002wh,Anastasiou:2002yz,Ravindran:2003um}
and the NLO differential cross section 
$d{\hat \sigma}_{ab}/d{\hat y} \,dq_T^2$ \cite{Glosser:2002gm,Ravindran:2002dc}
at large $q_T$.
To perform our NNLO calculation, we take advantage of these available results:
both observables are indeed computed up to relative order $\as^2$ 
with respect to the Born level cross section $\sigma_H^{(0)}(\as)$. 
We rewrite the $q_T$ integration in Eq.~(\ref{inte}) as follows:
\begin{align}
\label{master}
\int_0^{Q_0^2}dq_T^2 
\;\f{d{\hat \sigma}_{ab}}{dq_T^2}(q_T,M,{\hat s};\as) 
&
\equiv \int_0^{+\infty} dq_T^2 
\;\f{d{\hat \sigma}_{ab}}{dq_T^2}(q_T,M,{\hat s};\as)
-\int_{Q_0^2}^{+\infty} dq_T^2 
\;\f{d{\hat \sigma}_{ab}}{dq_T^2}(q_T,M,{\hat s};\as)
\nn \\
&=
{\hat \sigma}_{ab}^{({\rm tot})}(M,{\hat s};\as)
-\int_{Q_0^2}^\infty dq_T^2 \int^{+\infty}_{-\infty} d{\hat y}
\;\f{d{\hat \sigma}_{ab}}{d{\hat y} \,dq_T^2}({\hat y},q_T,M,{\hat s};\as) 
\;\;.
\end{align}
The cumulative partonic cross section over the range $0 < q_T < Q_0$ is thus
obtained by subtraction according to Eq.~(\ref{master}): 
we start from the total cross 
section\footnote{We actually use the expressions of 
Ref.~\cite{Ravindran:2003um},
which are given for general colour factors, $C_F=(N_c^2-1)/(2N_c)$ and 
$C_A=N_c$, of $SU(N_c)$.} 
${\hat \sigma}_{ab}^{({\rm tot})}$ and we subtract the contribution due
to the $q_T$ cross section in the `large'-$q_T$ region where $q_T > Q_0$
(in the context of Eq.~(\ref{master}), `large' values of $q_T$ generically means
`non-vanishing' values of $q_T$). The differential cross section 
$d{\hat \sigma}_{ab}/d{\hat y}\, d q_T^2$ in the integrand on the right-hand side
of Eq.~(\ref{master}) is presented in Ref.~\cite{Glosser:2002gm} in complete
analytic form: we use this form and we explicitly carried out the integrations
over ${\hat y}$ and $q_T$.
Since $q_T > Q_0$, these integrations can directly be performed in four
space-time dimensions, with no further use of dimensional 
regularization.
At NLO, the cumulative partonic cross section can be computed in explicit
analytic form for arbitrary values of $Q_0$, and the analytic result 
\cite{Catani:2001cr} is recalled below.
%At NLO, the cumulative partonic cross section can be computed in explicit
%analytic form\footnote{The NLO analytic result was already
%reported in Sect.~4 of Ref.~\cite{Catani:2001cr}. Indeed, at the NLO,
%the cumulative cross section of Eq.~(\ref{inte}) exactly coincides with the
%jet-vetoed cross section $\sigma^{\rm veto}$ of Ref.~\cite{Catani:2001cr},
%provided we identify $Q_0=p_T^{\rm veto}$, where $p_T^{\rm veto}$ is 
%the jet veto parameter.}
%for arbitrary values of $Q_0$.
At the NNLO, we limit ourselves to analytically computing
the cumulative cross section in the limit $Q_0 \ll M$, thus neglecting the terms
of ${\cal O}(Q_0^2/M^2)$ on the right-hand side 
of Eqs.~(\ref{eqr2s}) or (\ref{eqr2}).

The NLO analytic result for the cumulative partonic cross section in
Eqs.~(\ref{inte})
and (\ref{master})
was already presented in Ref.~\cite{Catani:2001cr}. 
Indeed, at the NLO,
the cumulative cross section exactly coincides with the
jet-vetoed cross section $\sigma^{\rm veto}$ in Sect.~4 of 
Ref.~\cite{Catani:2001cr},
provided we identify $Q_0=p_T^{\rm veto}$, where $p_T^{\rm veto}$ is 
the jet veto parameter. Considering arbitrary values of $Q_0$, the NLO function
${\hat R}^{(1)}_{ab}(z,M/Q_0)$ of Eqs.~(\ref{inte}) and (\ref{eqnR})
has the following form \cite{Catani:2001cr}:
\begin{equation}
\label{coeffveto}
{\hat R}^{(1)}_{ab}(z,M/Q_0) = G_{ab}^{{\rm veto} (1)}(z;\pitcut) = 
G_{ab}^{(1) \,({\rm tot})}(z) - \Delta G_{ab}^{(1)}(z;\pitcut) \;
\Theta\!\left(1-\pitcut\right)\;\;,
\end{equation}
where the variable $\pitcut$ depends on $z$ and $Q_0/M$,
\begin{equation}
\label{ptoverm}
\pitcut =
\pitcut(z,Q_0/M) \equiv \frac{2Q_0 {\sqrt z}}{(1-z)M}
\;\;.
\end{equation}
The two terms on the right-hand side of Eq.~(\ref{coeffveto}) are in one-to-one
correspondence with the NLO contribution to the two terms on the 
right-hand side of Eq.~(\ref{master}). The partonic functions 
$\Delta G_{ab}^{(1)}(z;\pitcut)$ are presented in Eq.~(20) of 
Ref.~\cite{Catani:2001cr}, and $G_{ab}^{(1) \,({\rm tot})}(z)$ 
are the partonic
functions of the NLO total cross section \cite{nlotot}
($G_{ab}^{(1) \,({\rm tot})}$ is denoted by $G_{ab}^{(1)}$ in 
Ref.~\cite{Catani:2001cr}, and we have introduced the superscript 
`$({\rm tot})$' to avoid confusion with the functions in our Eq.~(\ref{gone})).
The explicit expressions of $G_{ab}^{(1) \,({\rm tot})}(z)$ can be found in
Eqs.~(2.7)--(2.9) of Ref.~\cite{Catani:2001ic}, 
which uses the same overall normalization
as in Eq.~(\ref{coeffveto}).

The NLO coefficient functions ${\hat R}^{(1;k)}_{ab}(z)$ $(k=0,1,2)$
of Eq.~(\ref{eqr1s}) are obtained by performing the logarithmic expansion of 
the right-hand side of Eq.~(\ref{coeffveto}) at small values of $Q_0$.
The limit $Q_0 \ll M$ of Eq.~(\ref{coeffveto}) is not completely 
straightforward, since the functions ${\hat R}^{(1;k)}_{ab}(z)$ contain
`plus'-distributions
%`+'-distributions 
of the variable $z$.
To illustrate this point we consider, for example, the complete expression of
$\Delta G_{gg}^{(1)}(z;\pitcut)$ in Eq.~(20) of 
Ref.~\cite{Catani:2001cr} and, neglecting terms that trivially vanish if $Q_0
\to 0$, we obtain
\beeq
\label{deltagg}
%\Delta G_{gg}^{(1)}(z;\pitcut) \!\!\!\!\!\!\!\!&&
&& \Delta G_{gg}^{(1)}(z;\pitcut) \;
\Theta\!\left(1-\pitcut\right) = - \,C_A \;\frac{11 \,(1-z)^3}{6 z} \nn \\
&& 
+ 2 \,{\hat P}_{gg}(z)  \;
  \ln \left[ \frac{(1-z) M}{2 {\sqrt z} \,Q_0}
\left( 1 + {\sqrt {1- \frac{4 Q_0^2}{M^2 (1-z)^2}}}\,
\right) \right] \;
\Theta\!\left(1-z -\frac{2Q_0}{M}\right)
 + {\cal O}\!\left(\frac{Q_0^2}{M^2}\right) \,,
\eeeq
where ${\hat P}_{gg}(z)$ (see Eq.~(21) in Ref.~\cite{Catani:2001cr}) is the 
customary LO Altarelli--Parisi splitting function.
The sole non-trivial point related to the limit $Q_0 \to 0$ of
Eq.~(\ref{deltagg}) is due to the fact that ${\hat P}_{gg}(z)$ is proportional
to $1/(1-z)$ and, thus, singular when $z \to 1$.
At finite values of $Q_0$ this singularity is screened by the $\Theta$-function
in Eq.~(\ref{deltagg}), and the limit $Q_0 \to 0$ has to be properly treated
by introducing customary (mathematical) distributions, such as
$\delta(1-z)$ and `$+$'-distributions (see, e.g., Eq.~(9) in 
Ref.~\cite{Catani:2001cr}), that act onto smooth functions defined over the
range $0 \leq z \leq 1$.
In the specific case of Eq.~(\ref{deltagg}), these distributions occur in the
following expressions:
\beq
\label{pluslim}
\frac{\ln^k (1-z)}{1-z} \;\Theta\!\left(1-z -\frac{2Q_0}{M}\right)
= \left[ \,\frac{\ln^k (1-z)}{1-z} \,\right]_+ \,+
\frac{(-1)^k}{k+1} \ln^{k+1}\left(\frac{M}{2Q_0}\right) \,\delta(1-z) + 
{\cal O}\!\left(\frac{Q_0}{M}\right) \,,
\eeq
\beq
\label{deltalim}
\frac{1}{1-z} \,\ln \left[ \frac{1}{2}
\left( 1 + {\sqrt {1- \frac{4 Q_0^2}{M^2 (1-z)^2}}}
\right) \right] 
\Theta\!\left(1-z -\frac{2Q_0}{M}\right) =
\left( - \frac{\pi^2}{24} + \frac{1}{2} \ln^22 \right) \,\delta(1-z) + 
{\cal O}\!\left(\frac{Q_0}{M}\right) \,.
\eeq
Inserting Eqs.~(\ref{pluslim}) and (\ref{deltalim}) in Eqs.~(\ref{deltagg})
and (\ref{coeffveto}), we obtain the logarithmic expansions in 
Eqs.~(\ref{eqr1s}) and (\ref{eqr1}),
and the reader can directly crosscheck the correct values of 
$\Sigma_{gg\ito gg}^{H(1;2)}(z), \Sigma_{gg\ito gg}^{H(1;1)}(z)$
(see Eqs.~(63) and (64) in Ref.~\cite{Bozzi:2005wk}) 
and ${\cal H}_{gg\ito gg}^{H(1)}$
(see Eq.~(84) in Ref.~\cite{Bozzi:2005wk}, or Eqs.~(15) and (18) herein).

Our explicit computation of the coefficient functions 
${\cal H}_{gg\ito ab}^{H(2)}$ and, more generally, of the NNLO function 
${\hat R}^{(2)}_{ab}(z,M/Q_0)$ in
Eq.~(\ref{eqr2}) closely follows the same steps that we have just 
illustrated at the NLO. 
The details are too complicated and lengthy to present here. 
Using Eq.~(\ref{master}), we obtain the NNLO analogue of Eq.~(\ref{coeffveto})
and the corresponding partonic function  $\Delta G_{ab}^{(2)}$ arises from the
$q_T$ (and $y$) integration of the last term on the right-hand side of
Eq.~(\ref{master}). We carry out this analytic integration by neglecting terms
that trivially vanish if $Q_0 \to 0$, and we obtain an NNLO analytic expression
that is (conceptually) analogous  to Eq.~(\ref{deltagg}).
The final step of the NNLO calculation is analogous to that in 
Eqs.~(\ref{pluslim}) and (\ref{deltalim}), 
and it involves a proper treatment of the limit $Q_0 \ll M$
for several functions that become singular at the endpoint $z=1$.

In this 
%letter 
paper we have considered the production of the SM Higgs boson at hadron 
colliders. We have presented the NNLO analytic calculation of the cross 
section at small values of $q_T$ 
(see Eqs.~(\ref{inte}) and (\ref{eqr2s})).
The NNLO result is compared 
(see Eq.~(\ref{eqr2}))
with the predictions 
%(see Eq.~(\ref{}))
of transverse-momentum resummation
(see Eqs.~(\ref{qtycrossgg}) and (\ref{whath})).
The comparison gives a crosscheck of the factorization 
formula (\ref{whath})
(see also Eq.~(\ref{qq2}))
and allows us to determine the previously unknown resummation coefficients 
at ${\cal O}(\as^2)$.
These are the coefficient functions ${\cal H}^{H(2)}_{gg\ito ab}(z)$
(see Eqs.~(\ref{h2qq})--(\ref{h2gg}))
and the related coefficients $C^{(2)}_{g\,q}$ and $C^{(2)}_{g\,g}$
(see Eqs.~(\ref{gq22}) and (\ref{gg22})), 
which control the dependence on the rapidity
of the Higgs boson.
These coefficients can be implemented in resummed calculations of the inclusive
$q_T$ distribution at full NNLL accuracy.
Using the method of Ref.~\cite{Catani:2007vq},
the same coefficients 
%can be 
%are used 
are necessary
to perform the fully-exclusive 
perturbative calculation up to NNLO.

\noindent {\bf Acknowledgements.}
%This work has been supported in part by the European Commission through 
%the 'LHCPhenoNet' Initial Training Network PITN-GA-2010-264564.
This work was supported in part by the Research Executive Agency (REA) 
of the European Union under the Grant Agreement number PITN-GA-2010-264564 \\
({\it LHCPhenoNet},  Initial Training Network).


\begin{thebibliography}{99}

%%\cite{Catani:2000jh}
%\bibitem{Catani:2000jh}
%See the list of references in Sect.~5 of S.~Catani {\it et al.},
%  %``QCD,''
%hep-ph/0005025, in Proceeding of the CERN Workshop on {\em Standard Model 
%%Physics (and more) at the LHC}, eds. G.~Altarelli and M.L.~Mangano 
%%(CERN 2000-04, Geneva, 2000), p.~1.
% %%CITATION = HEP-PH/0005025;%%

%\cite{Dokshitzer:hw}
\bibitem{Dokshitzer:hw}
%\item \label{Dokshitzer:hw}
Y.~L.~Dokshitzer, D.~Diakonov and S.~I.~Troian,
%``On The Transverse Momentum Distribution Of Massive Lepton Pairs,''
Phys.\ Lett.\  B {\bf 79} (1978) 269,
%%CITATION = PHLTA,B79,269;%%
%``Hard Processes In Quantum Chromodynamics,''
Phys.\ Rep.\  {\bf 58} (1980) 269;
%%CITATION = PRPLC,58,269;%%
%\cite{Parisi:1979se}
%\bibitem{Parisi:1979se}
G.~Parisi and R.~Petronzio,
%``Small Transverse Momentum Distributions In Hard Processes,''
Nucl.\ Phys.\ B {\bf 154} (1979) 427;
%%CITATION = NUPHA,B154,427;%%
%\cite{Curci:1979bg}
%\bibitem{Curci:1979bg}
G.~Curci, M.~Greco and Y.~Srivastava,
%``QCD Jets From Coherent States,''
Nucl.\ Phys.\ B {\bf 159} (1979) 451.
%%CITATION = NUPHA,B159,451;%%

%\cite{Collins:1981uk}
\bibitem{Collins:1981uk}
J.~C.~Collins and D.~E.~Soper,
%``Back-To-Back Jets In QCD,''
Nucl.\ Phys.\ B {\bf 193} (1981) 381
[Erratum-ibid.\ B {\bf 213} (1983) 545],
%%CITATION = NUPHA,B193,381;%%
%\cite{Collins:va}
%\bibitem{Collins:va}
%J.~C.~Collins and D.~E.~Soper,
%``Back-To-Back Jets: Fourier Transform From B To K-Transverse,''
Nucl.\ Phys.\ B {\bf 197} (1982) 446.
%%CITATION = NUPHA,B197,446;%%

%\cite{Collins:1984kg}
\bibitem{Collins:1984kg}
J.~C.~Collins, D.~E.~Soper and G.~Sterman,
%``Transverse Momentum Distribution In Drell-Yan Pair And W And Z Boson 
%Production,''
Nucl.\ Phys.\ B {\bf 250} (1985) 199.
%%CITATION = NUPHA,B250,199;%%

%\cite{Catani:2000vq}
\bibitem{Catani:2000vq}
  S.~Catani, D.~de Florian and M.~Grazzini,
  %``Universality of non-leading logarithmic contributions in transverse
  %momentum distributions,''
  Nucl.\ Phys.\  B {\bf 596} (2001) 299.
  %[arXiv:hep-ph/0008184].
  %%CITATION = NUPHA,B596,299;%%

%\cite{Catani:2010pd}
\bibitem{Catani:2010pd}
  S.~Catani and M.~Grazzini,
  %``QCD transverse-momentum resummation in gluon fusion processes,''
  Nucl.\ Phys.\  B {\bf 845} (2011) 297.
  %[arXiv:1011.3918 [hep-ph]].
  %%CITATION = NUPHA,B845,297;%%

%\cite{Kodaira:1981nh}
\bibitem{Kodaira:1981nh}
J.~Kodaira and L.~Trentadue,
%``Summing Soft Emission In QCD,'
Phys.\ Lett.\ B {\bf 112} (1982) 66,
%%CITATION = PHLTA,B112,66;%%
%\ref{Kodaira:1982cr}
%\item \label{Kodaira:1982cr}
%J.~Kodaira and L.~Trentadue,
%``Soft Gluon Effects In Perturbative Quantum Chromodynamics,''
report SLAC-PUB-2934 (1982),
%\cite{Kodaira:1982az}
%\bibitem{Kodaira:1982az}
%J.~Kodaira and L.~Trentadue,
%``Single Logarithm Effects In Electron - Positron Annihilation,''
Phys.\ Lett.\ B {\bf 123} (1983) 335.
%%CITATION = PHLTA,B123,335;%%


\bibitem{Davies:1984hs}
%\cite{Davies:1984hs}
%\bibitem{Davies:1984hs}
C.~T.~H.~Davies and W.~J.~Stirling,
%``Nonleading Corrections To The Drell-Yan Cross-Section At Small Transverse
%Momentum,''
Nucl.\ Phys.\  B {\bf 244} (1984) 337;
%%CITATION = NUPHA,B244,337;%%
%\bibitem{Davies:1984sp}
%\cite{Davies:1984sp}
%\bibitem{Davies:1984sp}
  C.~T.~H.~Davies, B.~R.~Webber and W.~J.~Stirling,
  %``Drell-Yan Cross-Sections At Small Transverse Momentum,''
  Nucl.\ Phys.\  B {\bf 256} (1985) 413.
  %%CITATION = NUPHA,B256,413;%%

%\cite{Catani:vd}
\bibitem{Catani:vd}
S.~Catani, E.~D'Emilio and L.~Trentadue,
%``The Gluon Form-Factor To Higher Orders: Gluon Gluon Annihilation 
%At Small Q-Transverse,''
Phys.\ Lett.\ B {\bf 211} (1988) 335.
%%CITATION = PHLTA,B211,335;%%

\bibitem{deFlorian:2001zd}
%\cite{deFlorian:2000pr}
%\bibitem{deFlorian:2000pr}
  D.~de Florian and M.~Grazzini,
  %``Next-to-next-to-leading logarithmic corrections at small transverse
  %momentum in hadronic collisions,''
  Phys.\ Rev.\ Lett.\  {\bf 85} (2000) 4678,
  %[arXiv:hep-ph/0008152].
  %%CITATION = PRLTA,85,4678;%%
%\cite{deFlorian:2001zd}
%\bibitem{deFlorian:2001zd}
  %D.~de Florian and M.~Grazzini,
  %``The structure of large logarithmic corrections at small transverse
  %momentum in hadronic collisions,''
  Nucl.\ Phys.\  B {\bf 616} (2001) 247.
  %[arXiv:hep-ph/0108273].
  %%CITATION = NUPHA,B616,247;%%

%\cite{Becher:2010tm}
\bibitem{Becher:2010tm}
  T.~Becher and M.~Neubert,
  %``Drell-Yan production at small q_T, generalized parton distributions and the
  %collinear anomaly,''
  report HD-THEP-10-13
  (arXiv:1007.4005 [hep-ph]).
  %%CITATION = ARXIV:1007.4005;%%

%\cite{Catani:2007vq}
\bibitem{Catani:2007vq}
  S.~Catani and M.~Grazzini,
  %``An NNLO subtraction formalism in hadron collisions and its application to
  %Higgs boson production at the LHC,''
  Phys.\ Rev.\ Lett.\  {\bf 98} (2007) 222002.
  %[arXiv:hep-ph/0703012].
  %%CITATION = PRLTA,98,222002;%%

%\cite{Catani:2009sm}
\bibitem{Catani:2009sm}
  S.~Catani, L.~Cieri, G.~Ferrera, D.~de Florian and M.~Grazzini,
  %``Vector boson production at hadron colliders: a fully exclusive QCD
  %calculation at NNLO,''
  Phys.\ Rev.\ Lett.\  {\bf 103} (2009) 082001.
  %[arXiv:0903.2120 [hep-ph]].
  %%CITATION = PRLTA,103,082001;%%

%\cite{Bozzi:2010xn}
\bibitem{Bozzi:2010xn}
  G.~Bozzi, S.~Catani, G.~Ferrera, D.~de Florian and M.~Grazzini,
  %``Production of Drell-Yan lepton pairs in hadron collisions:
  %Transverse-momentum resummation at next-to-next-to-leading logarithmic
  %accuracy,''
  Phys.\ Lett.\  B {\bf 696} (2011) 207.
  %[arXiv:1007.2351 [hep-ph]].
  %%CITATION = PHLTA,B696,207;%%



% higgs qt res.



%\cite{Balazs:2000wv}
\bibitem{Balazs:2000wv}
  C.~Balazs and C.~P.~Yuan,
  %``Higgs boson production at the LHC with soft gluon effects,''
  Phys.\ Lett.\  B {\bf 478} (2000) 192;
  %[arXiv:hep-ph/0001103].
  %%CITATION = PHLTA,B478,192;%%
%\cite{Cao:2009md}
%\bibitem{Cao:2009md}
  Q.~H.~Cao, C.~R.~Chen, C.~Schmidt and C.~P.~Yuan,
  %``Improved Predictions for Higgs Q(T) at the Tevatron and the LHC,''
  report ANL-HEP-PR-09-20 (arXiv:0909.2305 [hep-ph]).
  %%CITATION = ARXIV:0909.2305;%%


%\cite{Berger:2002ut}
\bibitem{Berger:2002ut}
  E.~L.~Berger and J.~w.~Qiu,
  %``Differential cross-section for Higgs boson production including all orders
  %soft gluon resummation,''
  Phys.\ Rev.\  D {\bf 67} (2003) 034026,
  %[arXiv:hep-ph/0210135].
  %%CITATION = PHRVA,D67,034026;%%
%\cite{Berger:2003pd}
%\bibitem{Berger:2003pd}
  %E.~L.~Berger and J.~w.~Qiu,
  %``Differential cross-sections for Higgs boson production at Tevatron collider
  %energies,''
  Phys.\ Rev.\ Lett.\  {\bf 91} (2003) 222003.
  %[arXiv:hep-ph/0304267].
  %%CITATION = PRLTA,91,222003;%%  

%\cite{Bozzi:2003jy}
\bibitem{Bozzi:2003jy}
  G.~Bozzi, S.~Catani, D.~de Florian and M.~Grazzini,
  %``The q(T) spectrum of the Higgs boson at the LHC in QCD perturbation
  %theory,''
  Phys.\ Lett.\  B {\bf 564} (2003) 65.
  %[arXiv:hep-ph/0302104].
  %%CITATION = PHLTA,B564,65;%%

%\cite{Kulesza:2003wi}
\bibitem{Kulesza:2003wi}
  A.~Kulesza and W.~J.~Stirling,
  %``Nonperturbative effects and the resummed Higgs transverse momentum
  %distribution at the LHC,''
  JHEP {\bf 0312} (2003) 056
  %[arXiv:hep-ph/0307208].
  %%CITATION = JHEPA,0312,056;%%  

%\cite{Kulesza:2003wn}
\bibitem{Kulesza:2003wn}
  A.~Kulesza, G.~F.~Sterman and W.~Vogelsang,
  %``Joint resummation for Higgs production,''
  Phys.\ Rev.\  D {\bf 69} (2004) 014012.
  %[arXiv:hep-ph/0309264].
  %%CITATION = PHRVA,D69,014012;%%

\bibitem{Bozzi:2005wk}
  G.~Bozzi, S.~Catani, D.~de Florian and M.~Grazzini,
  %``Transverse-momentum resummation and the spectrum of the Higgs boson at the
  %LHC,''
  Nucl.\ Phys.\ B {\bf 737} (2006) 73.
%  [arXiv:hep-ph/0508068].
%%CITATION = HEP-PH 0508068;%%

%\cite{Bozzi:2007pn}
\bibitem{Bozzi:2007pn}
  G.~Bozzi, S.~Catani, D.~de Florian and M.~Grazzini,
  %``Higgs boson production at the LHC: transverse-momentum resummation and
  %rapidity dependence,''
  Nucl.\ Phys.\  B {\bf 791} (2008) 1.
  %[arXiv:0705.3887 [hep-ph]].
  %%CITATION = NUPHA,B791,1;%%

%\cite{Mantry:2010mk}
\bibitem{Mantry:2010mk}
  S.~Mantry and F.~Petriello,
  %``Transverse Momentum Distributions from Effective Field Theory with
  %Numerical Results,''
  Phys.\ Rev.\  D {\bf 83} (2011) 053007.
  %[arXiv:1007.3773 [hep-ph]].
  %%CITATION = PHRVA,D83,053007;%%

%\cite{deFlorian:2011xf}
\bibitem{deFlorian:2011xf}
  D.~de Florian, G.~Ferrera, M.~Grazzini and D.~Tommasini,
  %``Transverse-momentum resummation: Higgs boson production at the 
  %Tevatron and the LHC,''
  JHEP {\bf 1111} (2011) 064.
  %[arXiv:1109.2109 [hep-ph]].
  %%CITATION = JHEPA,1111,064;%%


%\cite{Grazzini:2008tf}
\bibitem{Grazzini:2008tf}
  M.~Grazzini,
  %``NNLO predictions for the Higgs boson signal in the H->WW->lnu lnu and
  %H->ZZ->4l decay channels,''
  JHEP {\bf 0802} (2008) 043.
  %[arXiv:0801.3232 [hep-ph]].
  %%CITATION = JHEPA,0802,043;%%


%\cite{Ellis:1975ap}
\bibitem{Ellis:1975ap}
  J.~R.~Ellis, M.~K.~Gaillard and D.~V.~Nanopoulos,
  %``A Phenomenological Profile of the Higgs Boson,''
  Nucl.\ Phys.\  B {\bf 106} (1976) 292;
  %%CITATION = NUPHA,B106,292;%%
%\cite{Shifman:1979eb}
%\bibitem{Shifman:1979eb}
  M.~A.~Shifman, A.~I.~Vainshtein, M.~B.~Voloshin and V.~I.~Zakharov,
  %``Low-Energy Theorems for Higgs Boson Couplings to Photons,''
  Sov.\ J.\ Nucl.\ Phys.\  {\bf 30} (1979) 711.
  %[Yad.\ Fiz.\  {\bf 30} (1979) 1368].
  %%CITATION = YAFIA,30,1368;%%

%\cite{Kramer:1996iq}
\bibitem{Kramer:1996iq}
  M.~Kramer, E.~Laenen and M.~Spira,
  %``Soft gluon radiation in Higgs boson production at the LHC,''
  Nucl.\ Phys.\  B {\bf 511} (1998) 523.
  %[arXiv:hep-ph/9611272].
  %%CITATION = NUPHA,B511,523;%%

%\cite{Chetyrkin:1997iv}
\bibitem{Chetyrkin:1997iv}
  K.~G.~Chetyrkin, B.~A.~Kniehl and M.~Steinhauser,
  %``Hadronic Higgs decay to order alpha(s)**4,''
  Phys.\ Rev.\ Lett.\  {\bf 79} (1997) 353,
  %[arXiv:hep-ph/9705240].
  %%CITATION = PRLTA,79,353;%%
%\cite{Chetyrkin:1997un}
%\bibitem{Chetyrkin:1997un}
%  K.~G.~Chetyrkin, B.~A.~Kniehl and M.~Steinhauser,
  %``Decoupling relations to O (alpha-s**3) and their connection to low-energy
  %theorems,''
  Nucl.\ Phys.\  B {\bf 510} (1998) 61.
  %[arXiv:hep-ph/9708255].
  %%CITATION = NUPHA,B510,61;%%

%\cite{Kauffman:1991cx}
\bibitem{Kauffman:1991cx}
  R.~P.~Kauffman,
  %``Higher Order Corrections To Higgs Boson P(T),''
  Phys.\ Rev.\  D {\bf 45} (1992) 1512;
  %%CITATION = PHRVA,D45,1512;%%
%\cite{Yuan:1991we}
%\bibitem{Yuan:1991we}
  C.~P.~Yuan,
  %``Kinematics of the Higgs boson at hadron colliders: 
  %NLO QCD gluon resummation,''
  Phys.\ Lett.\  {\bf B283 } (1992)  395.


%%% NNLO sigmatot

%\cite{Harlander:2002wh}
\bibitem{Harlander:2002wh}
  R.~V.~Harlander and W.~B.~Kilgore,
  %``Next-to-next-to-leading order Higgs production at hadron colliders,''
  Phys.\ Rev.\ Lett.\  {\bf 88} (2002) 201801.
  %[arXiv:hep-ph/0201206].
  %%CITATION = PRLTA,88,201801;%%

%\cite{Anastasiou:2002yz}
\bibitem{Anastasiou:2002yz}
  C.~Anastasiou and K.~Melnikov,
  %``Higgs boson production at hadron colliders in NNLO QCD,''
  Nucl.\ Phys.\  B {\bf 646} (2002) 220.
  %[arXiv:hep-ph/0207004].
  %%CITATION = NUPHA,B646,220;%%



%\cite{Ravindran:2003um}
\bibitem{Ravindran:2003um}
  V.~Ravindran, J.~Smith and W.~L.~van Neerven,
  %``NNLO corrections to the total cross section for Higgs boson production  in
  %hadron hadron collisions,''
  Nucl.\ Phys.\  B {\bf 665} (2003) 325,
  %[arXiv:hep-ph/0302135].
  %%CITATION = NUPHA,B665,325;%%
%two-loop colour factors:
%\cite{Ravindran:2004mb}
%\bibitem{Ravindran:2004mb}
%  V.~Ravindran, J.~Smith, W.~L.~van Neerven,
  %``Two-loop corrections to Higgs boson production,''
  Nucl.\ Phys.\  {\bf B704 } (2005) 332.
  %[hep-ph/0408315].

%\cite{Glosser:2002gm}
\bibitem{Glosser:2002gm}
  C.~J.~Glosser and C.~R.~Schmidt,
  %``Next-to-leading corrections to the Higgs boson transverse momentum spectrum
  %in gluon fusion,''
  JHEP {\bf 0212} (2002) 016.
  %[arXiv:hep-ph/0209248].
  %%CITATION = JHEPA,0212,016;%%

%\cite{Ravindran:2002dc}
\bibitem{Ravindran:2002dc}
  V.~Ravindran, J.~Smith and W.~L.~Van Neerven,
  %``Next-to-leading order QCD corrections to differential distributions of
  %Higgs boson production in hadron hadron collisions,''
  Nucl.\ Phys.\  B {\bf 634} (2002) 247.
  %[arXiv:hep-ph/0201114].
  %%CITATION = NUPHA,B634,247;%%


%\cite{Catani:2001cr}
\bibitem{Catani:2001cr}
  S.~Catani, D.~de Florian and M.~Grazzini,
  %``Direct Higgs production and jet veto at the Tevatron and the LHC in NNLO
  %QCD,''
  JHEP {\bf 0201} (2002) 015.
  %[arXiv:hep-ph/0111164].
  %%CITATION = JHEPA,0201,015;%%


%\cite{Dawson:1991zj}
\bibitem{nlotot}
  S.~Dawson,
  %``Radiative corrections to Higgs boson production,''
  Nucl.\ Phys.\ B {\bf 359} (1991) 283;
  %%CITATION = NUPHA,B359,283;%%
  %\bibitem{Djouadi:1991tk}
  A.~Djouadi, M.~Spira and P.~M.~Zerwas,
  %``Production of Higgs bosons in proton colliders: QCD corrections,''
  Phys.\ Lett.\ {\bf B264} (1991) 440.
  %%CITATION = PHLTA,B264,440;%%

%\cite{Catani:2001ic}
\bibitem{Catani:2001ic}
  S.~Catani, D.~de Florian and M.~Grazzini,
  %``Higgs production in hadron collisions: Soft and virtual QCD corrections  at NNLO,''
  JHEP {\bf 0105} (2001) 025.
  %[hep-ph/0102227].
  %%CITATION = HEP-PH 0102227;%%
  
\end{thebibliography}
\end{document}